# An exceptionally bright flare from SGR 1806−20 and the origins of short-duration γ-ray bursts

K. Hurley[1], S.E. Boggs[1,2], D. M. Smith[3], R. C. Duncan[4], R. Lin[1], A. Zoglauer[1], S. Krucker[1], G. Hurford[1], H. Hudson[1], C. Wigger[5], W. Hajdas[5], C. Thompson[6], I. Mitrofanov[7], A. Sanin[7], W. Boynton[8], C. Fellows[8], A. von Kienlin[9], G. Lichti[9], A. Rau[9] & T. Cline[10]

[1]*UC Berkeley Space Sciences Laboratory, Berkeley, California 94720-7450, USA*

[2]*University of California, Department of Physics, Berkeley, California 94720, USA*

[3]*Physics Department and Santa Cruz Institute for Particle Physics, University of California, Santa Cruz, Santa Cruz, California 95064, USA*

[4]*University of Texas, Department of Astronomy, Austin, Texas 78712, USA*

[5]*Paul Scherrer Institute, 5232 Villigen PSI, Switzerland*

[6]*Canadian Institute of Theoretical Astrophysics, 60 St George Street, Toronto, Ontario M5S 3H8, Canada*

[7]*Space Research Institute (IKI), GSP7, Moscow 117997, Russia*

[8]*University of Arizona, Department of Planetary Sciences, Tucson, Arizona 85721, USA*

[9]*Max-Planck-Institut für extraterrestrische Physik, Giessenbachstrasse (Postfach 1312), 85748 (85741) Garching, Germany*

[10]*NASA Goddard Space Flight Center, Code 661, Greenbelt, Maryland 20771, USA*

**Soft-γ-ray repeaters (SGRs) are galactic X-ray stars that emit numerous short-duration (about 0.1 s) bursts of hard X-rays during sporadic active periods. They are thought to be magnetars: strongly magnetized neutron stars with emissions powered by the dissipation of magnetic energy. Here we report the detection of a long (380 s) giant flare from SGR 1806−20, which was much more luminous than any previous transient event observed in our Galaxy. (In the first 0.2 s, the flare released as much energy as the Sun radiates in a quarter of a million years.) Its power can be explained by a catastrophic instability involving global crust failure and magnetic reconnection on a magnetar, with possible large-scale untwisting of magnetic field lines outside the star. From a great distance this event would appear to be a short-duration, hard-spectrum cosmic γ-ray burst. At least a significant fraction of the mysterious short-duration γ-ray bursts therefore may come from extragalactic magnetars.**

In the magnetar model, SGRs are isolated neutron stars with teragauss exterior magnetic fields[1–4] and even stronger fields within[5,6], making them the most strongly-magnetized objects in the Universe. Four SGRs are known. Three of them have now emitted giant flares[7,8]. These exceptionally energetic outbursts begin with a brief (~ 0.2 s) spike of γ-rays with energies up to several MeV, containing most of the flare energy. The spikes are followed by tails lasting minutes, during which hard-X-ray emissions gradually fade while oscillating at the rotation period of the neutron star.

The first-known giant flare, observed on 5 March 1979, came from SGR 0525−66 in the Large Magellanic Cloud. Its fluence implied an energy >~$6\times10^{44}$ erg (ref. 9). The second-known giant flare came from an SGR in our Galaxy, SGR 1900+14, on 27 August 1998. Its energy, in hard X-rays and γ−rays, was ~$2\times10^{44}$ erg (refs 8, 10). Here we describe a third giant flare, which came from the galactic SGR 1806−20 on 27 December 2004. Particle and γ-ray detectors onboard the Reuven Ramaty High Energy Solar Spectroscopic Imager (RHESSI), and particle detectors aboard the Wind spacecraft, indicate that this event was ~100 times more energetic than the 27 August flare. Its initial γ-ray spike had a quasi-blackbody spectrum, characteristic of a relativistic pair/photon outflow with an energetically small contamination of baryons. This is consistent with the catastrophic release of (nearly) pure magnetic energy from a magnetar[3]. The tremendous luminosity of the initial spike means that similar events could be detected from distant galaxies. This could account for some, and perhaps all, of the mysterious short-duration, hard-spectrum cosmic γ-ray bursts (GRBs).

**The giant flare from SGR 1806–20**

On 27 December 2004, the International Gamma-Ray Astrophysics Laboratory[11] (INTEGRAL) reported the detection of a spectacular flare. Four other missions in the third interplanetary network of GRB detectors (the High Energy Neutron Detector and Gamma Sensor Head aboard Mars Odyssey[12], the solar-pointing RHESSI[13], particle and γ-ray detectors aboard Wind[14], and NASA's recently launched GRB observatory Swift[15]) also reported this event. The light curve is shown in Fig. 1. Triangulation constrains the flare position to a portion of an annulus consistent with SGR 1806–20's position (annulus centre J2000, right ascension 15 h 56 m 37 s, declination −20° 13′ 50″, annulus radius 30.887±0.030°). No other known or candidate SGR lies within this area of the sky. SGR 1806–20 was 5.25° from the Sun at the time of these observations.

A ~1-s-long precursor was observed 142 s before the flare, with a roughly flat-topped profile (Fig. 1 inset). Its spectrum can be fitted with an optically thin thermal bremsstrahlung function with $kT\approx15$ keV. The precursor's >3-keV fluence was $1.8\times10^{-4}$ erg cm$^{-2}$, implying an energy of $4.8\times10^{42}d_{15}^2$ erg, where $d_{15} = (d / 15$ kpc), and $d$ is the distance to SGR 1806-20. Note that $0.8<d_{15}<1$ is likely for SGR 1806−20, owing to the apparent association of the SGR with a compact (~10 arcsec) stellar cluster[16,17]. The large energy and unusual light curve of the precursor distinguish it from most common SGR bursts. This and its proximity in time to the giant flare suggest that it is causally related.

The initial spike of the giant flare lasted for ~0.2 s. Its rise and fall times were $\tau_{rise}\leq 1$ ms and $\tau_{decay}\approx 65$ ms, similar to those of the other giant flares[8,18]. The spike's intensity drove all X- and γ-ray detectors into saturation, but particle detectors aboard RHESSI and Wind made reliable measurements. (The Supplementary Information describes our extensive Monte Carlo simulations of these particle detectors and has a full discussion of systematic uncertainties.) The RHESSI particle detector data imply a spike fluence in photons >30 keV of $(1.36\pm0.35)$ erg cm$^{-2}$, making this the most intense cosmic or solar transient ever observed (in terms of photon energy flux at Earth). The time-resolved energy spectrum, as measured by the Wind particle detectors, is consistent with a cooling blackbody (Fig. 2) with average temperature $T_{spike}=(175\pm25)$ keV. The spike energy is thus $E_{spike}=(3.7\pm0.9)\times10^{46}d_{15}^2$ erg, assuming isotropic emission. The peak flux in the first 0.125 s was $L_{spike}=2\times10^{47}d_{15}^2$ erg s$^{-1}$. Evidently, this event briefly outshone all the stars in the Galaxy put together by a factor of ~$10^3$.

The spike was followed by a hard-X-ray tail modulated with a period of 7.56 s, detected by the RHESSI γ-ray detectors, which were by this time unsaturated, for 380 s. This period agrees with the neutron star rotation period as inferred from cyclic modulations of its quiescent soft-X-ray counterpart[2]. The fluence in 3–100-keV photons during the tail phase is $4.6\times10^{-3}$ erg cm$^{-2}$ or $E_{tail}\approx 1.2\times10^{44}d_{15}^2$ erg.

**Physical interpretation**

This event can be understood as a result of a catastrophic instability in a magnetar. Strong shearing of the neutron star's magnetic field, combined with growing thermal pressure, appears to have forced an opening of the field outward, launching a hot fireball. The release of energy above a rate of ~$10^{42}$ erg s$^{-1}$ (less than one part in $10^4$ of the peak flare luminosity) into the magnetosphere leads to the formation of a hot, thermal pair plasma ($kT\approx 0.1$–1 MeV)[5]. The fast initial rise $\tau_{rise}\leq 1$ ms is consistent with a magnetospheric instability with characteristic time $\tau_{mag}\approx(R/0.1V_A)\approx 0.3$ ms, where $R\approx 10$ km and $V_A\approx c$ is the Alfven velocity in the magnetosphere, and $c$ is the speed of light[3]. This process must have occurred repeatedly, given that the hard initial spike persisted for a duration ~$10^3\tau_{mag}$. Indeed, there is evidence for spike variability in this and other giant flares[8,19,20]. The resulting outflow emitted a quasi-blackbody spectrum as it became optically thin, with spectral temperature comparable to the temperature at its base, because declining temperature in the outflow is compensated by the relativistic blueshift[21]. For luminosity $L_{spike}=10^{47}L_{47}$ erg s$^{-1}$ emerging from a zone with radius $R\approx 10$ km, the expected spectral temperature is $T_{spike}=(L_{spike}/4\pi acR^2)^{0.25}=200\,L_{47}^{0.25}$ keV, neglecting complications of magnetospheric stresses and intermittency. Almost all the pairs annihilated, and the outflow was only weakly polluted by baryons, as is clear from the extended, weak radio afterglow that followed the flare[22]. Note that we do not expect significant beaming of such powerful emissions from such a slowly rotating star.

When the outflow ceased, a trapped fireball was evidently left behind: an optically thick photon-pair plasma confined by closed field lines near the star[3,23]. The luminosity and lifetime of the tail (see the fitted curve in Fig. 3) are consistent with a fireball cooling rate that is limited by the transparency of the surface layers, where the temperature is ~10 keV and the plasma is dominated by ions and electrons[3,23,24]. The condition that the magnetic field must be strong enough to confine energy $E_{tail}$ within a distance $\Delta R \approx 10$ km of the star yields a rough bound on the dipole field, $B_{dipole} > 2\times 10^{14}(\Delta R/10\text{ km})^{-3/2}\,[(1+\Delta R/R)/2]^3$ G, similar to bounds implied by the previous giant flares[3,8].

A clue to the nature of the instability comes from the spike's ~0.2-s duration, which is similar to the durations of other giant flare spikes[7,8,18] and of most other SGR bursts[25]. In the magnetar model, SGR activity results from the unwinding of a strong, toroidal magnetic field inside the star, and the transfer of magnetic helicity across the surface[23,26]. Such a twist propagates along the poloidal magnetic field $B_p=10^{15}B_{p15}$ G with a speed $V_A \approx B_p/(4\pi\rho)^{-0.5}$ that is weakly dependent on the twist amplitude. The time to cross the neutron star interior (density $10^{15}\,\rho_{15}$ g cm$^{-3}$) is $\Delta t \approx 2R/V_A \approx 0.2 B_{p15}^{-1}$ s.

Thus the 27 December event could have been a crustal instability that drove helicity from the star[23,26]. The unwinding of a toroidal magnetic field embedded in the crust is strongly impeded by the stable stratification and near-incompressibility of the crust[23]. Because of the energetic cost of forming isolated dislocation surfaces that cross the magnetic flux surfaces, the crust must undergo smooth and vertically differential torsional motion when it fails, which requires a fundamental breakdown of its solid structure. The maximum field energy which can be released is estimated by balancing elastic and magnetic stresses in the crust: $E_{max} \sim 1\times 10^{46}\,(\theta_{max}/10^{-2})^2\,B_{P15}^{-2}$ erg, where $\theta_{max}$ is the yield strain. Supplying the energy of the December 27 flare thus requires a relatively large yield strain, as well as a large twist of the crust with angular displacement approaching $\sim 0.5\,B_{P15}^{-1}$ radian.

Since March 2004, SGR 1806−20 has been very burst-active[27], while its quiescent X-ray brightness has increased by a factor of 2 to 3, and its spectrum has hardened dramatically[28]. Evidently, crust failure has enhanced the twist in the external magnetic field, with growing magnetospheric currents[26]. The free energy of such an exterior magnetic twist can reach a modest fraction ($\sim 10^{-1}$) of the untwisted exterior dipole field energy, $E_{\text{twist}} \approx 10^{-2} B_{\text{dip}}^2 R^3 \approx 10^{46} B_{\text{p15}}^2$ erg, with more energy in the non-potential components of higher multipoles. Some of this energy could be catastrophically released via reconnective simplification of the magnetosphere[26,29]. An extreme possibility, consistent with the flare energy, is a global magnetospheric untwisting. This would predict a dramatic post-flare drop in the stellar spin-down rate, as well as greatly diminished, softened and less strongly pulsed X-ray emissions. However, a pure magnetospheric instability would proceed much faster than ~0.2 s. Note also that the detection of accelerated spin-down[30] several months after previous active periods of SGRs 1806−20 and 1900+14 betrays a net increase in the magnetospheric twist during the X-ray bursts, and in the 27 August 1998 giant flare. Observations of SGR 1806-20's spin-down over the coming year will provide important constraints on the location of the non-potential magnetic field that was dissipated during the flare.

**Short-duration GRBs and magnetars**

If observed from a great distance, only the brief, initial hard spike of the 27 December flare would be evident. Thus distant extragalactic magnetar flares (MFs) would resemble the mysterious short-duration GRBs[31,32]. These hard-spectrum events have long been recognized as a separate class of GRBs[33–37] but have never been identified with any counterparts[38].

The Burst and Transient Source Experiment (BATSE) on the Compton Gamma-Ray Observatory was a landmark experiment of the 1990s that produced a catalogue[39] of more than 2,000 GRBs. How many of these bursts were MFs? Taking the 27 December event as our prototype and adopting the 50% trigger-efficiency flux[40] of 0.5 photons cm$^{-2}$ s$^{-1}$ for the 256-ms timescale yields a BATSE sampling depth of $D_{\text{BATSE}} = 30\, d_{15}$ Mpc. If such events generally happen once every $\tau = 30$ yr in galaxies like the Milky Way (such as has now occurred in the Milky Way itself) then the BATSE detection rate of MFs is $\dot{N}(BATSE) = 19 d_{15}^3 (\tau/30\,\text{yr})^{-1}\,\text{yr}^{-1}$. Here we have estimated the effective number of galaxies like the Milky Way within $D_{\text{BATSE}}$ of Earth by multiplying the local blue luminosity density[41] $j_b = 5.8 \times 10^{41} h_{70}$ erg Mpc$^{-3}$ by the sampling volume $(4\pi/3) D_{\text{BATSE}}^3$, and dividing by the blue luminosity of the Milky Way as estimated in the Supplementary Information. We use blue emissions as a benchmark because SGRs are Population I objects, the post-supernova remnants of massive, short-lived, blue stars. Thus, over 9.5 yr of operation with half-sky coverage, BATSE probably detected $180 d_{15}^3 (\tau/30\,\text{yr})^{-1}$ MFs, representing $0.4 d_{15}^3 (\tau/30\,\text{yr})^{-1}$ of all BATSE short-duration bursts. There is evidence of 100-s-long soft tails in the co-added time histories of many short-duration BATSE GRBs[42,43]; but not in any single event. For the brightest observed BATSE short-duration, hard-spectrum GRB (trigger number 6293), we find that the ratio of the tail-to-peak fluence is <0.5%, compared to our measured ratio for the 27 December event of 0.34%. Thus BATSE was not sensitive enough to have detected MF tails in single bursts.

The GRB observatory Swift[44] was designed, in part, to unravel the short-duration GRB mystery. How many MFs will Swift spot? The Swift Burst Alert Telescope has a photon flux sensitivity (50–300 keV) that is ~5 times better than BATSE[45], corresponding to a trigger threshold of ~0.10 photons cm$^{-2}$ s$^{-1}$. Thus for our prototype MF, $D_{\text{Swift}}$=70 $d_{15}$ Mpc. The expected rate of MF detections, given Swift's sky coverage of 1.4 steradians, is then $\dot{N}(Swift) = 53 d_{15}^3 (\tau/30\,\text{yr})^{-1}$ yr$^{-1}$, or about one MF per week. Of course, the galactic rate of MFs, $\Gamma = \tau^{-1}$, is very uncertain. Given that there has occurred one MF with peak luminosity in the range $10^{47}$ erg s$^{-1}$ in our Galaxy during $t_0 = 30$ yr of observations, the bayesian probability distribution for the underlying galactic rate $\Gamma$ of such bright MFs is $(dP/d\Gamma)=t_0\exp(-\Gamma t_0)$, with expected value $\langle \Gamma \rangle = t_0^{-1}$. This implies that the probability that Swift will detect one or more MF per month is 80% for $d_{15} = 1$. The probabilities of detecting one or more event per {3, 6, 12, 24} months are {93, 96, 98, 99}%, respectively. Even if $d = 10$ kpc, the probabilities would be {78, 88, 94, 97}%. The prospects for observing MFs during Swift's 24-month prime mission are excellent.

Of course, all of the above estimates idealize MFs as 'standard candles' defined by the 27 December prototype. The actual luminosity function of MFs is unknown. It is possible that some MFs are significantly brighter than the 27 December event. For example, a magnetic instability on a rare magnetar with $B_{\text{dipole}} \approx 10^{16}$ G could release $10^{48}$ erg, and be detected by Swift out to ~1 Gpc. Nevertheless, we suspect that MFs constitute only a substantial subset of BATSE Class II GRBs, not all of them. The 175-keV blackbody spectrum would probably result in a significantly higher hardness ratio than that of the average short-duration burst[37]. The fact that Class II GRBs have $\langle V/V_{\text{max}} \rangle < 0.5$ does not seem consistent with all these events being local[23]. Moreover, no galaxies at $D$<100 Mpc were found for the Interplanetary Network positions of four short-duration GRBs[38].

**Studying extragalactic magnetars**

Swift can identify MFs via their positional correlations with galaxies, allowing the source distances from Earth to be inferred. A spiral galaxy of size ~30 kpc at distance $D_{Swift}$ spans ~3.4 arcmin, comparable to the Swift BAT location accuracy of $\Delta\theta_{BAT}\approx1$–4 arcmin. This localization can be greatly improved, to an accuracy of $\lesssim$~10 arcsec, if the oscillating tail of the flare is detected by Swift's X-ray Telescope (XRT) when it slews to observe the burst site within about 1 min. Our measurements of soft X-ray emissions in the giant flare tail (Fig. 4) make it possible to assess the prospects of XRT acquisition for the first time. Extrapolating our X-ray spectral fits down to 0.3 keV, we find that the 27 December pulsating tail produced a 0.3–10-keV incident fluence of $(0.18$–$1.6)\times10^{-3}$ erg cm$^{-2}$. The threshold fluence for XRT detection[44] is $2\times10^{-10}$ erg cm$^{-2}$, so that the 27 December flare tail could be marginally detected to a distance of $D_{tail}=10$–$40\ d_{15}$ Mpc. Thus only the nearest fraction $(D_{tail}/D_{Swift})^3\approx0.2$ of all MFs spotted by Swift will have detectable tails. We have verified that the soft X-rays are strongly pulsed (Fig. 5). For events within about 8 Mpc, simulations indicate that the magnetar's rotation period can be reliably determined. For more distant sources, the spectrum and the rapid flux decay will distinguish magnetar tail emissions from cosmic GRB afterglows.

The prospects of detecting extragalactic MFs with the Swift Ultra-Violet and Optical Telescope (UVOT) or ground-based optical telescopes are not wholly bleak. The trapped fireball is too tiny to emit detectably in this waveband. However, we can scale directly from the observed radio afterglow[22], which had spectral index $\alpha = -0.7$ and time decay $t^{-1.5}$ in the optically thin regime. Extrapolating to $10^{14.5}$ Hz gives $L_{opt}\approx4\times10^{37}t_3^{-1.5}$ erg s$^{-1}$ at a time $10^3\,t_3$ s post-flare. Such a source would have a brightness of 22nd magnitude at 1 Mpc for $t_3\approx1$.

Prospects are even better for the detection of X-ray afterglows[32]. SGR 1900+14 emitted strong nonthermal X-rays in the aftermath of the 27 August 1998 event[46], thought to be due to a heated magnetar crust[47]. If afterglow energy scales linearly with flare energy, as found in less energetic events[48], then a MF like the 27 December event would glow brighter by a factor of $f\approx100$, suggesting $L_X\approx2\times10^{39}(f/100)(t/1\text{ h})^{-0.7}$ erg s$^{-1}$. This could be detected by the Chandra X-ray Telescope within $D_{Chandra}\approx30(f/100)^{0.5}(\Delta t_{obs}/10^4\text{ s})^{0.5}(t/10\text{ h})^{-0.35}$ Mpc in an observation of duration $\Delta t_{obs}\ll t$ in seconds.

**New horizons and speculations**

The detection of extragalactic magnetars, if achieved by *Swift*, will open up a new field of astronomy. A catalogue of giant flare spikes, once assembled, will contain a wealth of information about magnetic instabilities in neutron stars. Information about the luminosity function of MFs, their range of durations, and possible spectral diversity (suggested by measurements of the 27 August event[8,49]; note that less compact flows than that of the 27 December event could show nonthermal spectra) will constrain magnetar physics and population diversity. Unusually bright flares may be detected from very young magnetars with rapid rotation periods and stronger fields than are observed in galactic SGRs. (The birthrate of SGRs is evidently so low that no stars younger than $\sim 10^3 - 10^4$ yr are observed in our galaxy.) MFs from very young magnetars may be disproportionately common in extragalactic studies because of their greater brightness and higher flare rate. More frequent cataclysms are expected in younger magnetars because magnetic diffusion slows down as stars age and cool[6].

We emphasize that most SGR activity is ultimately powered by the strong toroidal interior field of a magnetar, $B_\phi$, which is the remnant of the rapid differential rotation which the neutron star experienced at birth[1,5]. The energy of this field, $E_\phi \sim (1/6) B_\phi^2 R^3 \sim 2 \times 10^{49} B_{\phi,16}^2$ erg, where $B_{\phi,16} = (B_\phi / 10^{16} \text{ G})$, can power many flares of $\sim 10^{46}$ erg over a star's lifetime. Magnetic helicity is gradually transported outward via ambipolar diffusion and Hall drift[6], winding up the field within the crust and outside the star, and leading to catastrophic instabilitities[23,26]. (Note, however, that the strong, internal toroidal field stabilizes a magnetar against catastrophic decay of the exterior dipole field; compare with refs 5 and 32.) Measurements of SGR 1806−20's spin-down over the coming year will reveal whether the exterior magnetic helicity increased or decreased during the 27 December event. SGR 1806−20 may come to resemble an anomalous X-ray pulsar, with a diminished spin-down rate and a softer X-ray spectrum. SGR 0526−66 developed these characteristics, indicating weak magnetospheric currents, after the giant flare of 5 March 1979 (ref. 50). Sporadic, short bursts were observed from SGR 0525−66 until 1983 (ref. 51, but the source has not been observed to burst since then, suggesting that sub-crust stresses were (at least temporarily) relieved in the giant flare. We speculate that SGR 0526−66 and now SGR 1806−20 may have entered the 'low' phase in a magnetar activity cycle that involves changes in the rate of expulsion of magnetic helicity out of the star.

## SUPPLEMENTARY INFORMATION

## RHESSI and Wind particle detector data analysis

During the intense initial spike, all X- and γ-ray detectors experienced some degree of saturation, making reliable reconstruction of the time history and energy spectrum difficult or impossible. Many small, thin silicon particle detectors, on the other hand, had very low cross-sections for X- and γ-ray interactions, and therefore did not saturate, even though they did respond strongly to the peak. We have therefore analysed the observations of the Wind 3D plasma and energetic particle experiments[52] and of the RHESSI particle monitor detector[53] with the GEANT3 and GEANT4 simulation codes to obtain information about the initial spike (specifically, the first $kT$ in Fig. 1b, and the spectrum, time history and $kT$ in Fig. 2). The RHESSI particle detector has an area of 25 mm$^2$ and is 960 μm thick. Wind has six double-ended solid-state telescopes (SSTs), five with two back-to-back 1.5-cm$^2$, 300-μm-thick silicon detectors (called O and F, with nine and seven PHA channels, respectively), and one SST with a third, 15-cm$^2$, 500-μm-thick detector (T) in between. The multi-channel analysers covered the 20 keV to 11 MeV range with various time resolutions between 12 and 96 s, while the RHESSI detector had two discriminators with 50- and 620-keV thresholds that were read out with 0.125-s resolution. In each case, the simulations included the matter surrounding the detectors, and attempted to reproduce the observed count rates with incoming power-law, thermal bremsstrahlung, and blackbody energy spectra. In all cases, the power law and bremsstrahlung spectra were strongly rejected by the Wind data ($\chi^2$=42 and 69 for 10 degrees of freedom), and only the blackbody provided an acceptable fit ($\chi^2$=10 for 10 degrees of freedom). These fits were performed for the Wind detectors with the highest statistics (F and O), because they gave the strongest restriction on the error bars for the blackbody temperature (175±25 keV). A systematic error of ±10% was assumed for the Wind simulations. This is a typical conservative estimate for simulations of this type; it includes uncertainties in the masses and compositions in the structure surrounding the detector, as well as uncertainties in the detector size, volume, and calibration. Fits including all the Wind detectors are also consistent with these results. An additional systematic uncertainty of ±15% was included for the RHESSI data, to include the effects of absorption in the spacecraft structure and interception of photons scattered off the Earth's atmosphere. Both these effects were modelled in GEANT3, with the prediction that 25% of the incoming photons are removed by the spacecraft-absorption process, and an approximately equal number are added by the photon-interception process, but at lower energies, tending to soften the overall spectrum. The observed RHESSI response is consistent with the blackbody fit.

For the peak, the sum of the count rates from the 12 Wind detectors reached 1,900 counts per second per detector. The Wind particle detectors have a 600-ns shaping time, which is fast enough that pulse pile-up in the detectors is negligible. However, the overall throughput of the system is determined by the sampling rate of the multiplexed analogue-to-digital converters, which is not well quantified at these data rates. Therefore the overall livetime of the detectors is uncertain, and the responses cannot be used to measure the fluence, even though they are well within the count rate range of measuring the spectral shape correctly. Thus Wind was used to measure the spectral shape during the spike, and the RHESSI particle detector was used to derive the normalization.

The RHESSI particle detector counted 3,008 counts in the peak 125 ms of the flare. Its saturation level is approximately $10^5$ counts in 125 ms. Thus pulse pile-up is negligible. Because this detector has only two channels, it cannot strongly constrain the spectral shape, although it can confirm or reject the spectral shapes found by the Wind detectors, and it can determine the normalization of the Wind spectra accurately. These data were used to produce the time history and *kT* in the inset to Fig. 2.

**RHESSI γ-ray detector data analysis**

The RHESSI γ-ray detectors are segmented Ge detectors which record the time and energy of each photon interaction >3 keV. They were unsaturated after the initial spike. However, there are two structures that can attenuate the incoming photons in the observations of the oscillatory phase described here. The first is a shutter that was automatically put into place over the front segments as a response to the high count rates, and remained there for the first 272 s in Fig. 1. The second is the imaging grid structure above the detectors, which affects both the front and rear segments. However, as the spacecraft rotates, a direct (unattenuated) path exists to some of the detectors for brief intervals twice per rotation period. We call these intervals 'snapshots'. To minimize the effects of attenuation in Fig. 1a, the inset to Fig. 1a and the black curve in Fig. 5a, we have used counts >20 keV. To eliminate these effects in Fig. 4 and Fig. 5a, we have used the snapshot data. We have also used these snapshots to obtain the spectral temperatures in Fig. 1b. We have used the on-axis (0°) RHESSI response matrices for this analysis, which should reproduce reasonable flux numbers and spectral distributions. With the current matrices we are unable to distinguish strongly between thermal bremsstrahlung and blackbody spectral fits for the tail, so we have included both in this paper. We anticipate that further spectral analysis including response matrices for this source location (under construction) should discriminate between these models.

**Detectability of magnetar flares by BATSE and Swift**

We estimated the BATSE sampling depth for MFs using our peak incident flux from this flare in the standard BATSE 50–300-keV energy range (determined from our best-fit RHESSI Particle Detector fluence and WIND spectral fit), over the BATSE trigger timescales of 64, 256 and 1,024 ms. We find the optimal BATSE trigger timescale to be 256 ms (BATSE's P256). Given the 50%-efficiency trigger flux for P256 (ref. 13) of 0.50 photons cm$^{-1}$ s$^{-1}$, we determine that this flare would have been detected by BATSE to a distance of 31 Mpc. As a check, we analysed the 50–300-keV fluence of all the BATSE short-duration, hard-spectrum GRBs with durations $T_{90}$=0.1–0.2 s, and found a threshold fluence of ~5×10$^{-8}$ erg cm$^{-2}$, corresponding to comparable detection distance. This is lower than the distance originally quoted in GCN 2936 (ref. 39) as a result of our spectral fits—the black-body fit is much harder than typical GRB spectra, resulting in lower photon fluxes in the 50–300-keV range than a typical short-duration, hard-spectrum GRB spectrum with comparable energy flux. To estimate the Swift BAT sensitivity, we used a P256 (50–300 keV) photon flux sensitivity 5 times better than BATSE's (see figure 9 in ref. 45), corresponding to ~0.10 photons cm$^{-2}$ s$^{-1}$, for a limiting detection distance for BAT of 70 Mpc. As a check, the advertised energy flux sensitivity of ~10$^{-8}$ erg cm$^{-2}$ s$^{-1}$ yields an even larger limiting distance.

To estimate the BATSE sensitivity to pulsating tails, we examined the strongest short-duration, hard-spectrum GRB seen by BATSE, trigger number 6293. This GRB had a duration $T_{90}$=0.192 s, and a total fluence of 4.30×10$^{-5}$ erg cm$^{-2}$, dominated by photons of >300 keV. Given the background count rate in the 400-s period after this burst, we estimate a 5$\sigma$ upper limit on a 20–100-keV tail fluence of 2×10$^{-7}$ erg cm$^{-2}$, setting the BATSE upper limit on the ratio of tail-to-peak fluence of 0.5%.

To estimate the Swift XRT sensitivity to the pulsating tails, we used the XRT response available in the HEASARC WebPIMMS package. We developed a model of the pulsating X-ray tail from our time-dependent thermal bremsstrahlung fits over the course of the 380-s tail, assuming the average 3–10-keV pulse shape. Folding the time-dependent model through the XRT response, and assuming an optimistic 20-s slew time, we estimate a marginal 0.3–10-keV detection of the soft tail at 10–40 Mpc for blackbody and bremsstrahlung spectra. As a check, the 27 December tail produced an incident 0.3–10-keV fluence of 0.18−1.6×10$^{-3}$ erg cm$^{-2}$. The quoted threshold flux for XRT detection is 2×10$^{-14}$ erg cm$^{-2}$ s$^{-1}$ for a 10$^4$-s observation, corresponding to a fluence threshold of 2×10$^{-10}$ erg cm$^{-2}$. Comparing this with our measured X-ray fluence yields a comparable detection distance. We also determined that the magnetar rotation period can be picked out of the XRT data by Fast Fourier Transforms out to distances of ~2–8.5 Mpc (it is clearly visible by eye out to ~1–4 Mpc).

**Rate of magnetar flares**

To estimate the rate of extragalactic magnetar flares, we needed to estimate the blue luminosity of the Milky Way, $L_{b, MW}$. The synthetic Galactic model of ref. 54, based upon Hipparcos data and recent large-scale surveys in the optical and infrared, implies a Galactic stellar thin disk mass $M_{MW}=2\times10^{10}M_{\odot}$, where $M_{\odot}$ is the mass of the Sun. We divided this by $M/L_b=1.4M_{\odot}/L_{\odot}$, which was found from the average of 30 Milky-Way-like galaxies of types Sb to Sc with luminosities of $5\times10^9 L_{\odot}$ to $5\times10^{10} L_{\odot}$ within the Nearby Field Galaxy Survey (S. Kannappan, personal communication).

**Supplementary Information** accompanies the paper on **www.nature.com/nature.**

**Acknowledgements** We are grateful to J. Scalo, E. Vishniac and S. Kannappan for discussions and expert help. In the US, this work was supported by NASA. The INTEGRAL mission is supported by the German government via the DLR agency.

**Competing interests statement.** The authors declare that they have no competing financial interests.

**Correspondence** and requests for materials should be addressed to K.H. (khurley@ssl.berkeley.edu).

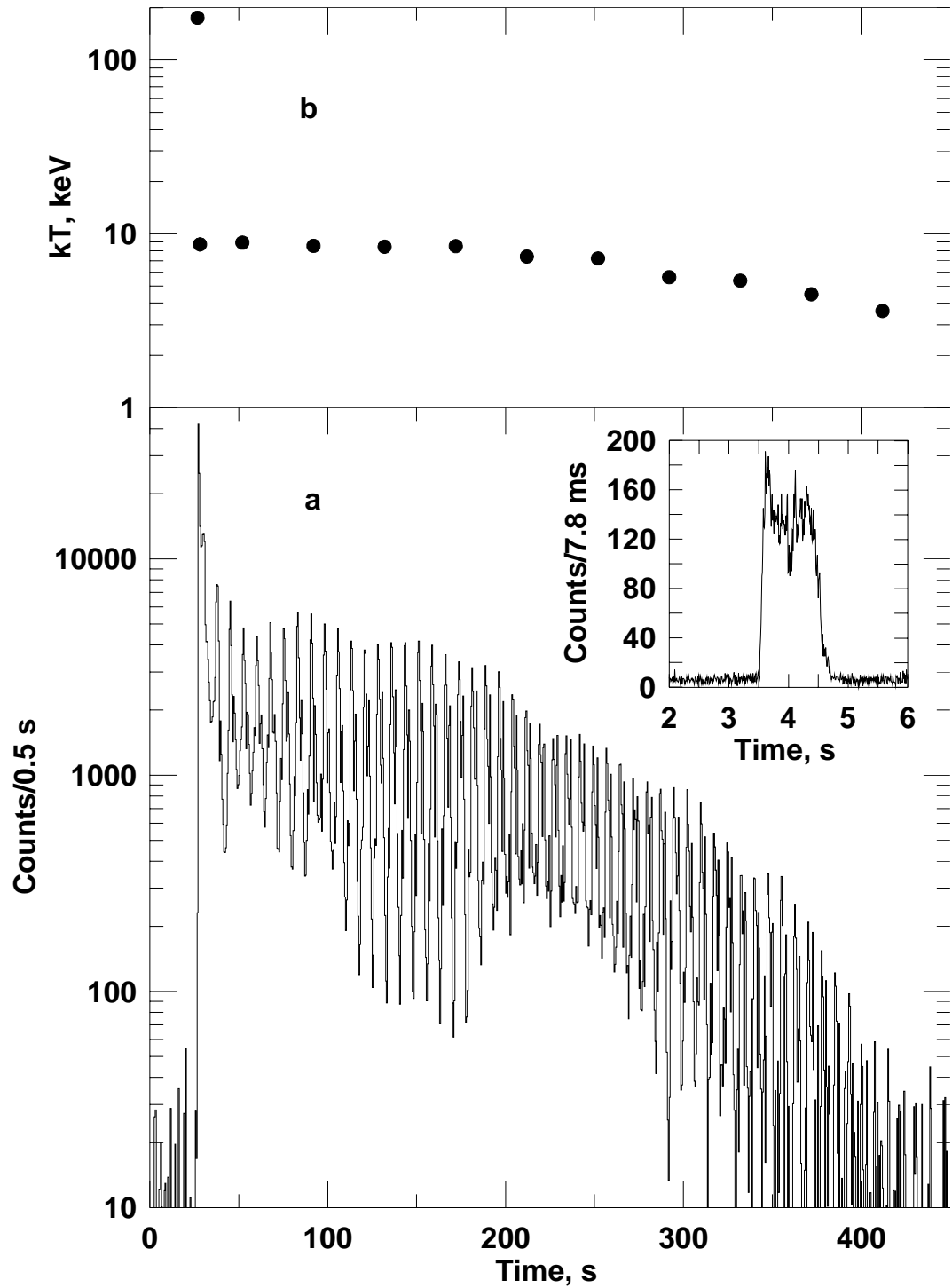

**Figure 1** Profiles of the 27 December 2004 giant flare. **a**, 20–100-keV time history plotted with 0.5-s resolution, from the RHESSI γ-ray detectors. Zero seconds

corresponds to 77,400 s Universal Time (UT). In this plot, the flare began with the spike at 26.64 s and saturated the detectors within 1 ms. The detectors emerged from saturation on the falling edge 200 ms later and remained unsaturated after that. Photons with energies >~20 keV are unattenuated; thus the amplitude variations in the oscillatory phase are real, and are not caused by any known instrumental effect (Supplementary Information). Inset, time history of the precursor with 8-ms resolution. Zero corresponds to 77,280 s UT. **b**, Spectral temperature versus time in the oscillatory phase. The temperature of the spike was determined by the RHESSI and Wind particle detectors; the temperatures of the oscillatory phase were measured by the RHESSI γ-ray detectors. Although RHESSI measured time- and energy-tagged photons >3 keV continuously, unattenuated spectra were measured for short 'snapshot' intervals only twice in each 4.06-s spacecraft spin period during the oscillatory phase (Supplementary Information). Preliminary spectral analysis (3–100 keV), using the RHESSI on-axis response matrices, is generally consistent with a single-temperature blackbody or optically thin thermal bremsstrahlung model; the blackbody temperatures have been plotted. The formal uncertainties in the oscillatory phase are smaller than the data points and are not shown.

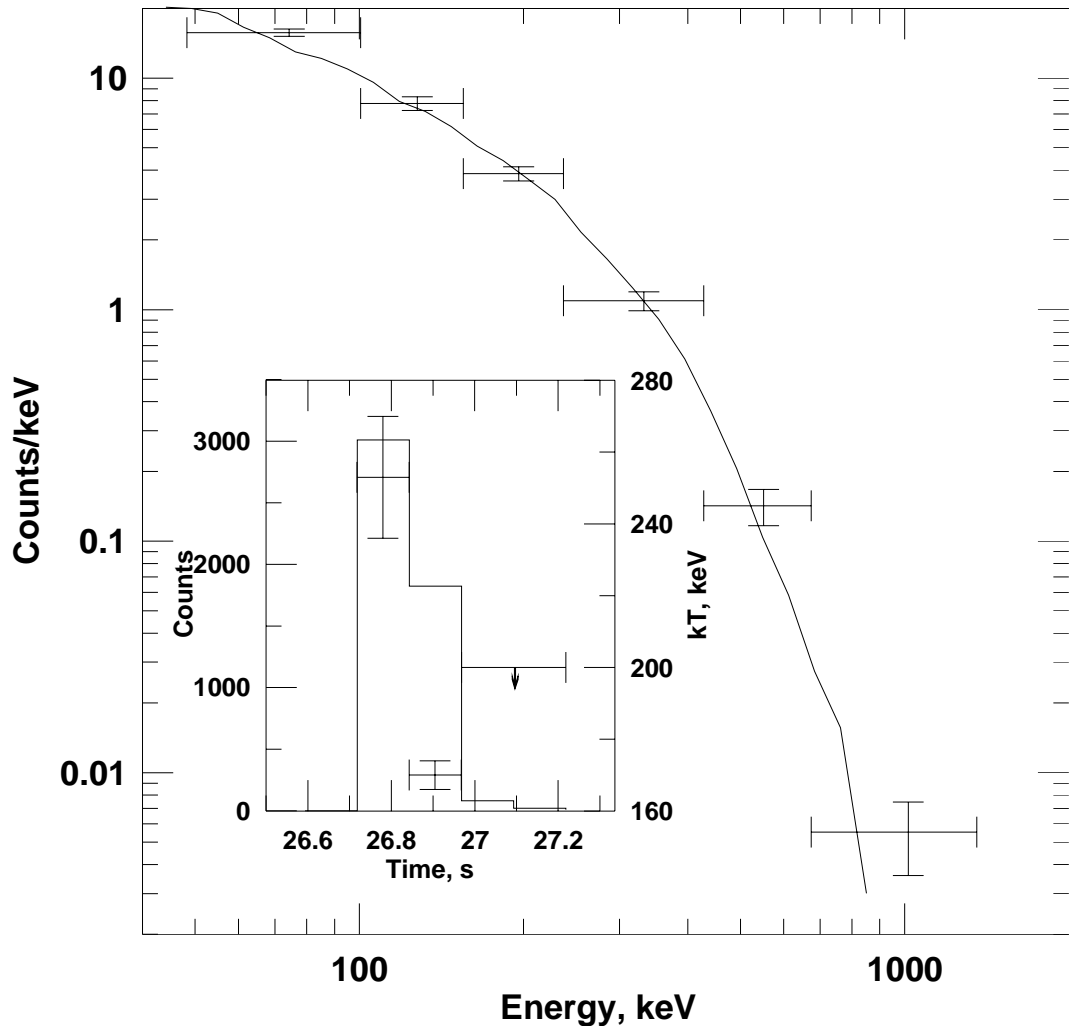

**Figure 2** Spectrum and time history of the initial spike, from the RHESSI and Wind particle detectors. The crosses show the spectrum measured by the Wind 3D O detector[52] with coarse time resolution that averages over the peak. The error bars are $1\sigma$, plus 10% systematic errors. The line is the best-fitting blackbody convolved with the detector response function; its temperature is 175±25 keV (Supplementary Information). Inset, the time history of the peak (histogram, left-hand scale) and of the blackbody temperature (error bars, right-hand scale) with 0.125-s resolution, from the RHESSI particle detector (ref. 35 and Supplementary Information). The error bars are $1\sigma$, plus 25% systematic errors.

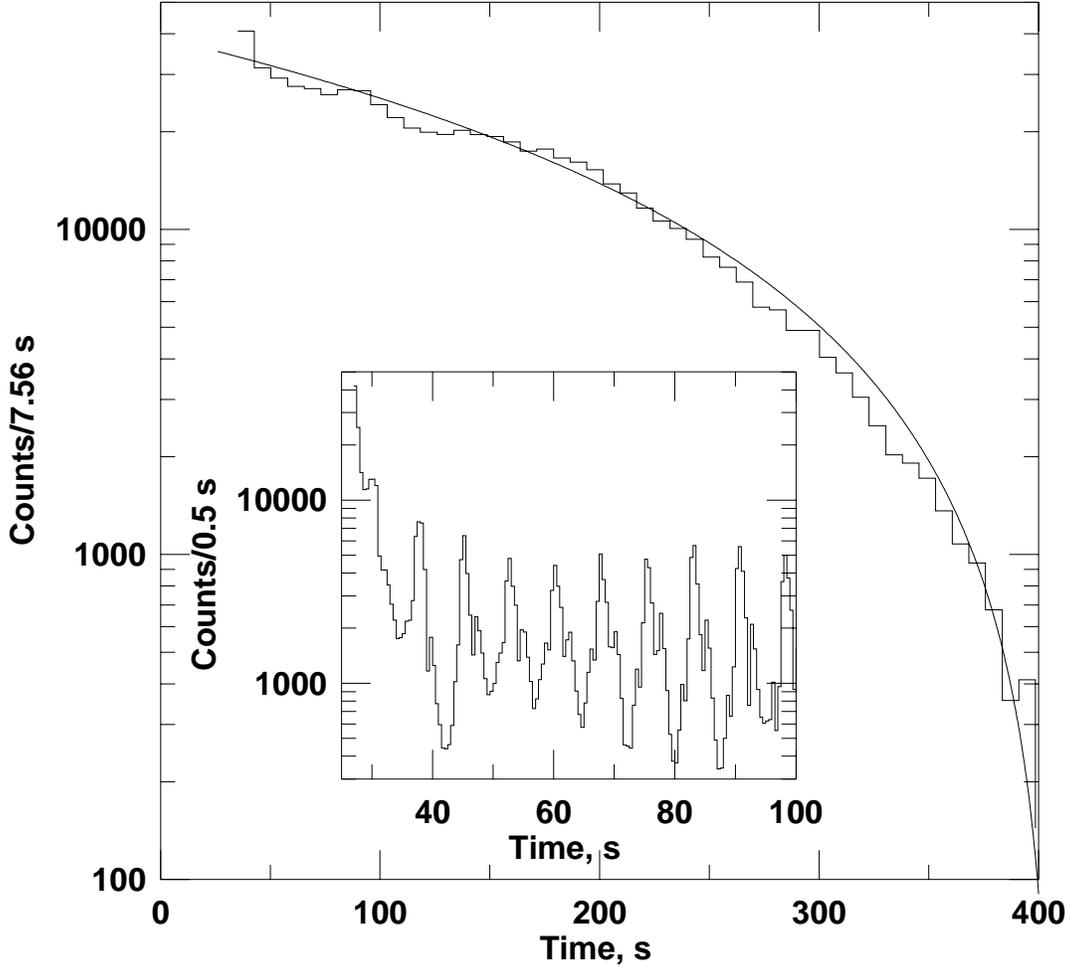

**Figure 3** Time-averaged counts in the tail phase of the giant flare, compared with the 'trapped fireball' model. Zero corresponds to 77,280 s UT. The step plot shows the RHESSI γ-ray detector data averaged over the 7.56-s rotation period of the neutron star. It is fitted by a simple model (smooth curve) that describes the emission from the cool surface of a magnetically confined plasma as it contracts and evaporates in a finite time: $L_x(t)=L_0[1-(t/t_{evap})]^{a/(1-a)}$ (ref. 49). We find $t_{evap}=382\pm3$ s, and the index $a = 0.606 \pm 0.003$ is near the value $a=2/3$ expected for a homogeneous, spherical trapped fireball[23,49]. Inset, RHESSI γ-ray detector light curve for the first ten cycles of the flare tail. The energy range is 20–100 keV. The first peak of the trapped fireball emission is evident on the falling edge of the hard spike at $t=30$ s. A changing two-peaked pulse–interpulse structure is present.

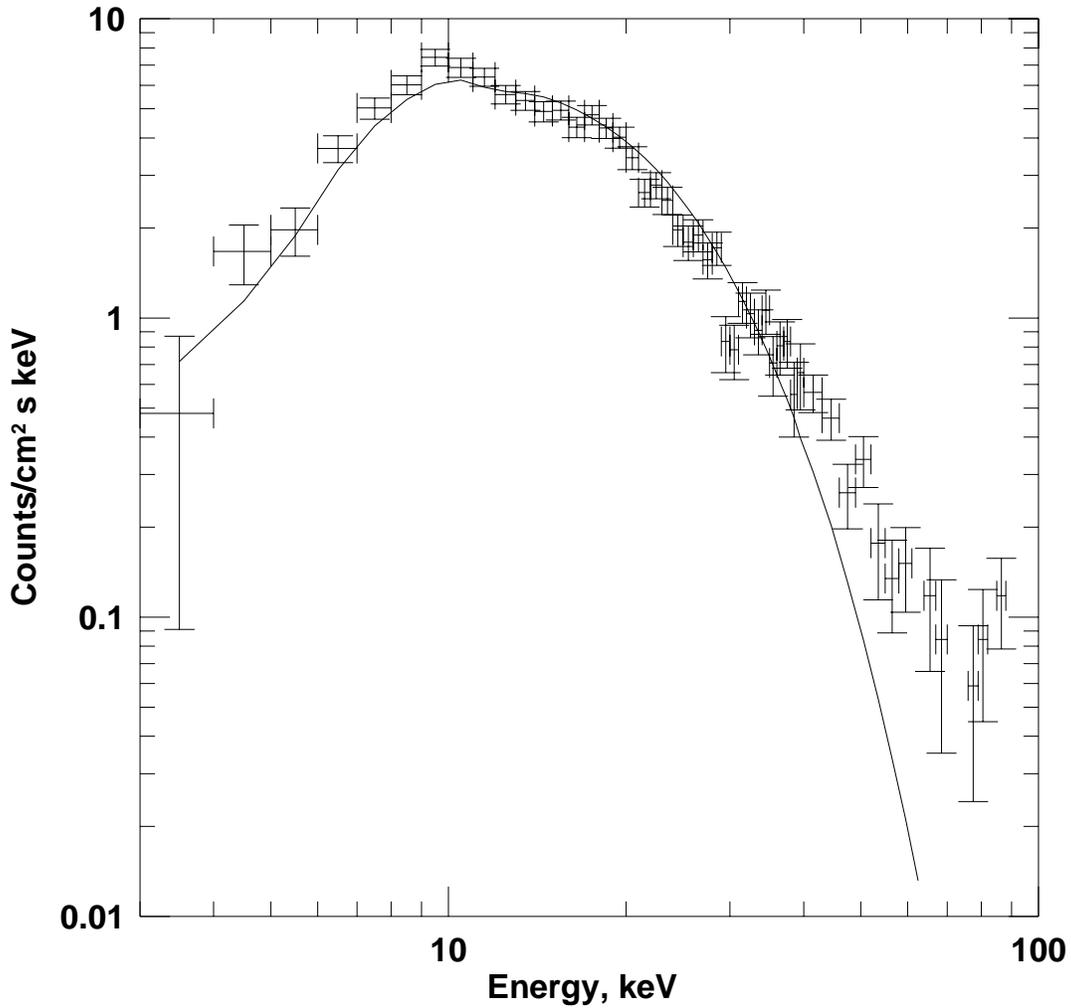

**Figure 4** 3–100-keV phase-averaged energy spectrum of the pulsed tail, from the RHESSI γ-ray detectors. The crosses show the measured spectrum with 1σ statistical error bars; the solid line represents a fit to a blackbody function $E^2(\exp(E/kT)-1)^{-1}$, where $E$ is the energy and $kT$=5.1±1.0 keV. This spectrum is averaged over various phases between 272 and 400 s in Fig. 1, corresponding to intervals where the photons could reach the detectors passing through a minimum amount of intervening materials (Supplementary Information). An optically thin thermal bremsstrahlung function with $kT \approx 22$ keV also provides a reasonable fit. The spectra show evidence of deviations from both models, probably due to the use of an approximate response matrix[24].

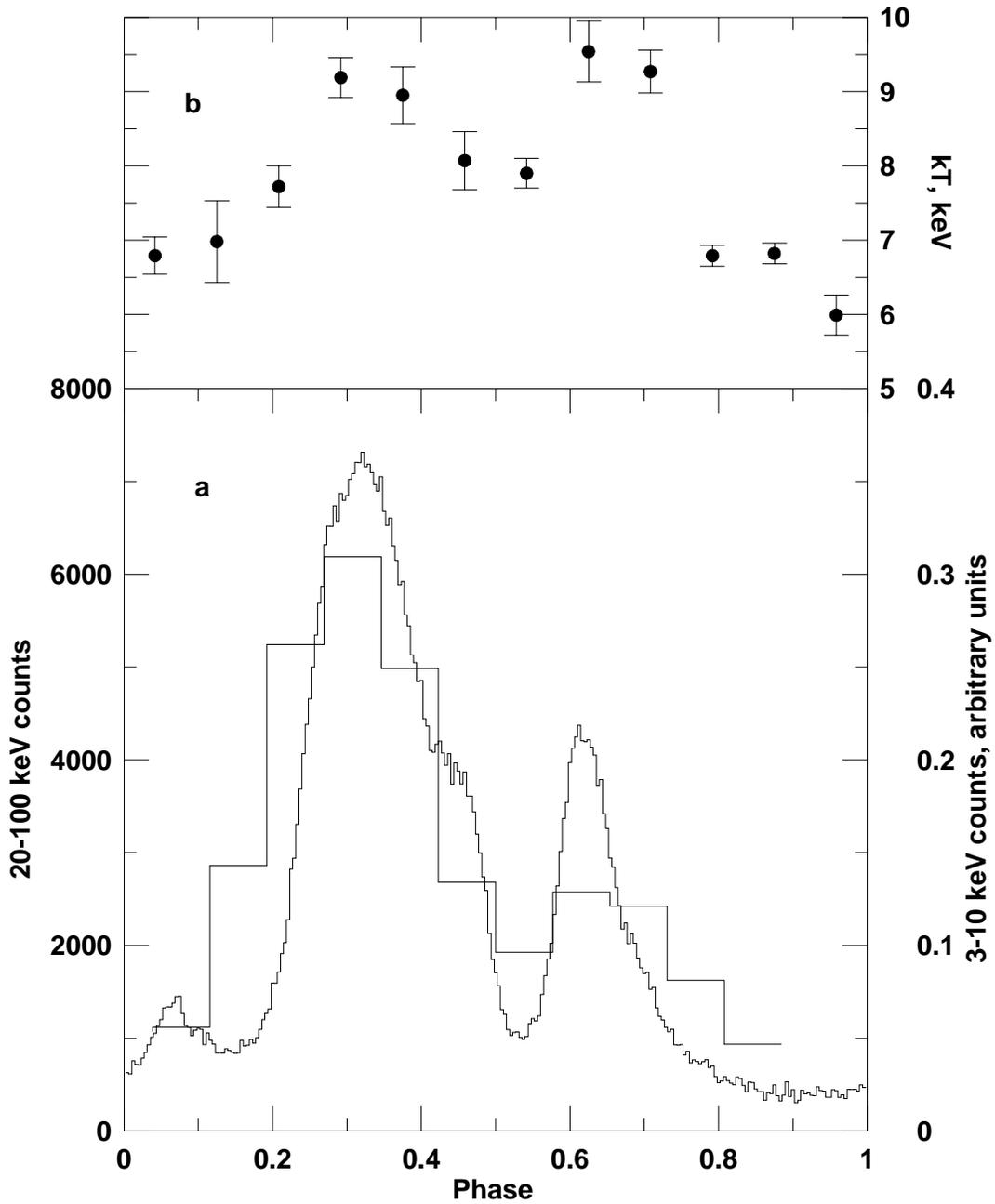

**Figure 5** Detailed profiles of the oscillations, from the RHESSI γ-ray detectors. **a**, RHESSI light curve for the oscillatory portion of the giant flare, folded modulo the 7.56-s neutron star rotation period (20–100 keV, fine resolution curve, and 3–10 keV, coarse resolution curve). **b**, The blackbody spectral temperature $kT$. The radius of the emitting surface varies between ~18 and 40 km at 15 kpc.